\documentclass[lettersize,journal]{IEEEtran}
\usepackage{amsmath,amsfonts}
\usepackage{algorithmic}
\usepackage{algorithm}
\usepackage{array}
\usepackage[caption=false,font=normalsize,labelfont=sf,textfont=sf]{subfig}
\usepackage{textcomp}
\usepackage{stfloats}
\usepackage{url}
\usepackage{verbatim}
\usepackage{graphicx}
\usepackage{cite}
\usepackage{siunitx}
\usepackage{color}
\newcommand{\revision}[1]{{\color{black}{#1}}}
\usepackage[normalem]{ulem}
\begin{document}

%%%%%Added formatting for ArXiv submission (comment out for final file submission)
\onecolumn
\textcopyright~2023 IEEE.  Personal use of this material is permitted.  Permission from IEEE must be obtained for all other uses, in any current or future media, including reprinting/republishing this material for advertising or promotional purposes, creating new collective works, for resale or redistribution to servers or lists, or reuse of any copyrighted component of this work in other works.
\newpage
\twocolumn

\title{Cognitive and Physical Activities\\ Impair Perception of Smartphone Vibrations} % this is just a temporary title!

\author{Kyle T.\ Yoshida\textsuperscript{$\star$},~\IEEEmembership{Student Member,~IEEE,} Joel X.\ Kiernan\textsuperscript{$\star$}, Rachel A.\ G.\ Adenekan,~\IEEEmembership{Student Member,~IEEE,} Steven H.\ Trinh, Alexis J.\ Lowber, Allison M.\ Okamura,~\IEEEmembership{Fellow,~IEEE,} and Cara M.\ Nunez,~\IEEEmembership{Member,~IEEE}

\vspace{-20pt}

        % <-this % stops a space
\thanks{Manuscript received XXXXXX XX, XXXX; revised XXXXXX XX, XXXX;
accepted XXXXXX XX, XXXX. Date of publication XXXXXX XX, XXXX; date of
current version XXXXXX XX, XXXX. This article was recommended for publication by Associate Editor XXXXXXXXXXXX and Editor-in-Chief XXXXXXXXXXXX upon evaluation of
the reviewers’ comments.} % <-this % stops a space
\thanks{K.\ T.\ Yoshida, J.\ X.\ Kiernan, R.\ A.\ G.\ Adenekan, S.\ H.\ Trinh, and A.\ M.\ Okamura are with the Mechanical Engineering Department and A.\ J.\ Lowber is with the Computer Science Department at Stanford University, Stanford, CA 94305.  (email: \{kyle3; joelx; adenekan; steventrinh; aokamura; lowbaj21\}@stanford.edu).} 
\thanks{C.\ M.\ Nunez is with the Sibley School of Mechanical and Aerospace Engineering at Cornell University, Ithaca, NY 14850. (email: cmn97@cornell.edu).}
\thanks{This work was supported in part by a grant from the Precision Health and Integrated Diagnostics Center at Stanford, the National Science Foundation Graduate Fellowship Program and grant 1830163, and the Stanford Graduate Fellowship Program.}
\thanks{\textsuperscript{$\star$}These authors contributed equally to this work.}}

\markboth{Journal of \LaTeX\ Class Files,~Vol.~XX, No.~X, XXXXXX~XXXX}%
{Shell \MakeLowercase{\textit{et al.}}: A Sample Article Using IEEEtran.cls for IEEE Journals}


\maketitle
\begin{abstract}
Vibration feedback is common in everyday devices, from virtual reality systems to smartphones. However, cognitive and physical activities may impede our ability to sense vibrations from devices. In this study, we develop and characterize a smartphone platform to investigate how a shape-memory task (cognitive activity) and walking (physical activity) impair human perception of smartphone vibrations. We measured how Apple's Core Haptics Framework parameters can be used for haptics research, namely how \textit{hapticIntensity} modulates amplitudes of 230~Hz vibrations. A 23-person user study found that physical ($p<0.001$) and cognitive ($p=0.004$) activity increase vibration perception thresholds. Cognitive activity also increases vibration response time ($p<0.001$). %\revision{\sout{Our results indicate that cognitive and physical activities impair vibration perception thresholds.}} 
This work also introduces a smartphone platform that can be used for out-of-lab vibration perception testing. Researchers can use our smartphone platform and results to design better haptic devices for diverse, unique populations.
\end{abstract}


\section{Introduction}
% Introduction outline (combining Kyle's comments and Cara's thoughts) - each number is a new paragraph
%1. talk about why studying distraction is important
%2. talk about prior work in the realm of distractions
%3. talk about why smartphone
%4. talk about iphone work - us and Hannah Stuart (could be combined with paragraph 3)
%5. give purpose of the paper and main contributions - should be written/structured something like below
%%We present our methods in developing a smartphone system and user study design in Section~\ref{sec:methods}, our preliminary results from a human-subjects user study in Section~\ref{sec:results}, and a forward-looking discussion in Section~\ref{sec:discussion}.

%  Haptics research and development traditionally takes place in a laboratory setting. However, translating this work to be successfully used out in-the-wild poses several challenges, many of them presented in~\cite{Blum2019OutsideLab}. An additiona
Real-life, out-of-lab haptic applications, including virtual reality, social interactions, wearables, and smartphones, are becoming increasingly common. However, the transition from in-lab to out-of-lab can present a variety of challenges for researchers~\cite{Blum2019OutsideLab}. Traditionally, haptics research studies are conducted in a highly controlled laboratory setting. This usually requires participants to remain stationary and focused on the provided haptic feedback. The use of haptic devices in the real-world, such as through smartphones, inherently involves a variety of cognitive and physical activities which may impede the perception of haptic cues, diminishing the performance of the haptic device~\cite{Blum2019OutsideLab}. A better understanding of how cognitive and physical distractions affect human perception could inform how laboratory studies and devices can be used in everyday applications. 

\begin{figure}[]
	\centering
    \includegraphics[width=0.7\columnwidth]{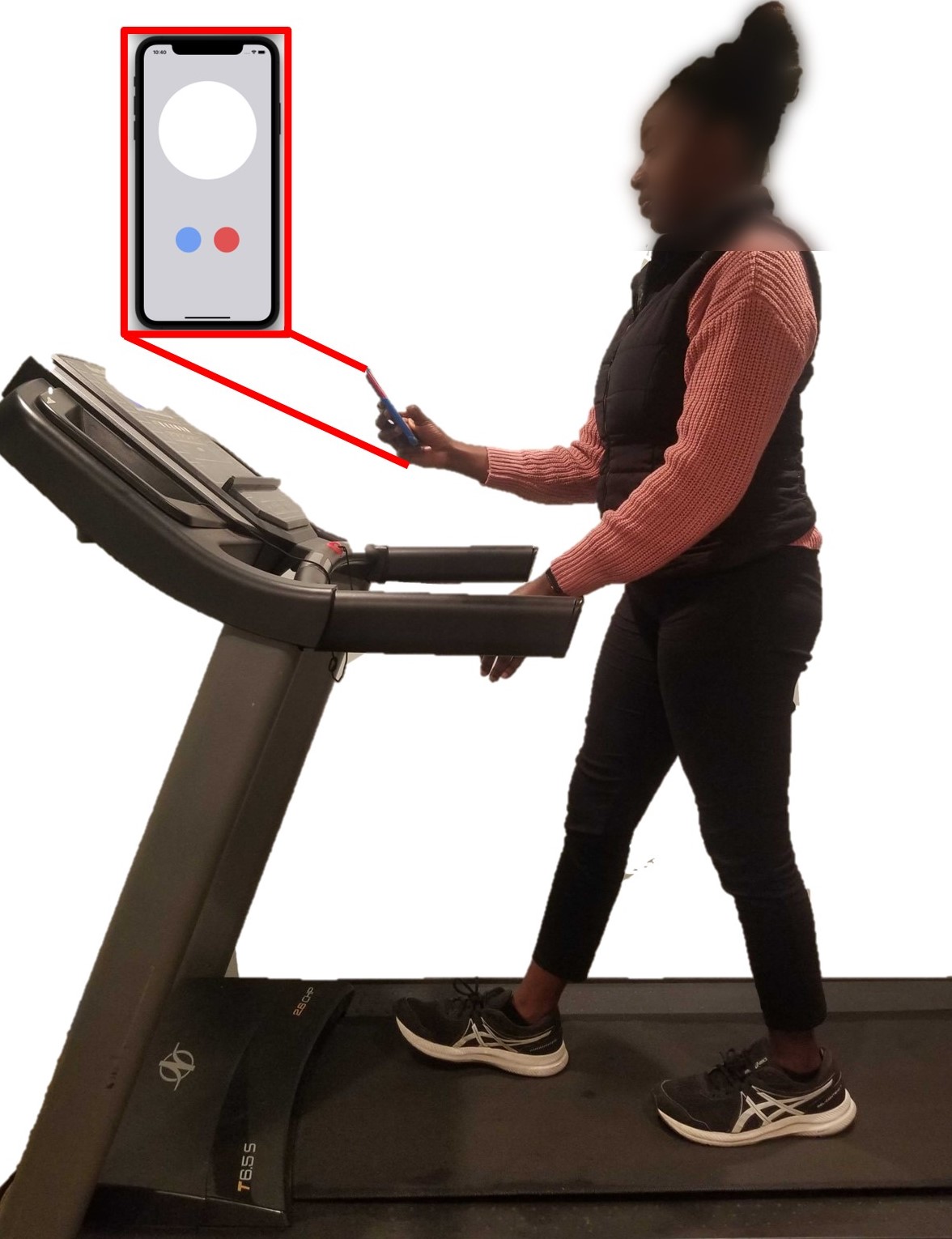}
	
	\caption{We developed an iOS app (inset) and case for an Apple iPhone 11 to determine how vibration perception changes with different levels of cognitive (shape-recall task or no task) and physical (walking or sitting) activity. Participants pressed a red button each time they felt a vibration and a blue button every time they saw a repeated shape. Participant responses were used to calculate vibration perception thresholds.}\label{fig:setup}
    \vspace{-12pt}
\end{figure}

\revision{Existing research has examined how haptic perception changes with physical or cognitive activity for specific lab-based systems. Prior work has shown that vibration perception decreases in the presence of elbow movement~\cite{post_zompa_chapman_1994}, wrist movement~\cite{yildiz_activeandpassive_2015}, finger movement~\cite{ANGEL1988cutaneoussensitivity}, and walking~\cite{karuei_acrossbodywalking}. Research has also found that walking on different textures~\cite{chapquouo2018restandwalking} and auditory distractions while moving~\cite{Chapwouo2018} also impact cutaneous perception at the feet.}

\revision{Some studies also evaluated the interplay between cognitive activity and cutaneous perception~\cite{Klatzky2007, Haghighi2020TheEO}, while others have used dual-task assessments to understand how to provide cutaneous feedback for driving~\cite{Ploch2017} and guidance~\cite{shah2018dualtaskreaching}. Cognitive activity also results in increased reaction time to roadway hazards~\cite{daddario_donmez_2019}. The combination of physical and cognitive activity also impacts cutaneous perception of vibration in distributed, full-body systems~\cite{karuei_acrossbodywalking}.}

%\revision{\sout{Research studying the effects of cognitive and physical activity on haptic perception is limited. Some studies do focus on evaluating the effects of distraction on various forms of human perception and attention. Chapwouo et al. found that auditory distractions inhibited subjects' ability to differentiate haptic ``messages" stimulated in their feet~\cite{Chapwouo2018}. Previous work has also found that haptic distractions impede the completion of cognitive tasks~\cite{Haghighi2020TheEO}. D`Addario et al. showed that cognitive distraction increases reaction time to roadway hazards~\cite{daddario_donmez_2019}.}}

In this work, we evaluate how distractions in the form of cognitive and physical activities impact vibration perception of stimuli from a smartphone. We target the universally-available smartphone because people frequently use their phones while multi-tasking in daily life. For example, one might receive and respond to notifications while walking to a location. Furthermore, we believe that results from the commercially-available smartphone would be useful for designing mobile-apps to account for distractions, while also informing the designs for custom haptic devices~\cite{yoshida_3dof_worldhaptics, socialtouch}. 

\begin{comment}
Developing and running a study on a mobile platform offers advantages over a strictly in-lab platform. First, a more diverse audience can be reached. Compared to research conducted at a typical research institution, a smartphone platform enables studies with people from diverse backgrounds outside of conventional lab settings~\cite{lee2018}. Second, experiments on a smartphone platform are more generalizable and on a common platform. In-lab setups are not always generalizable across different platforms due to the nature and application of haptic feedback~\cite{Chapwouo2018,Haghighi2020TheEO,oldmen_reactiontime}, but a smartphone-based study removes the reliance on in-lab platforms.
\end{comment}

Other researchers have also used smartphone vibrations to study perception for clinical tests. For instance, Adenekan et al. found that a staircase method could be used to measure vibration perception~\cite{adenekan_phone}, and Torres et al. found that smartphone vibration perception was correlated to perception from a monofilament exam~\cite{Torres2022}.  

In this work, we aim to understand the effects of both cognitive and physical activity on vibrotactile haptic perception through stimuli of varying intensities provided by a smartphone. \revision{Our work differs from prior studies because we use a widely-used smartphone platform to understand the effect of both cognitive and physical activities on perception threshold and response time in situations similar to that which one might encounter in daily life.} In Section~\ref{sec: smartphone platform}, we present the software and hardware elements of our smartphone system. We then describe our user study design and our results from a human-subjects user study in Section~\ref{sec: user study}, followed by a discussion of those results in Section~\ref{sec:discussion}. We conclude with future work in Section~\ref{sec:conclusion}.

\section{Smartphone Platform}
\label{sec: smartphone platform}

In this section, we describe the design of the smartphone platform. First, we discuss the development of a smartphone app that delivers a vibration-response task and shape-recall task. Next, we introduce a phone case used to normalize finger placement while participants hold the phone. Last, we present the characterization of the vibrations occurring at each finger location. This platform was then used for the user study presented in Section~\ref{sec: user study} to test how vibration perception changes with cognitive and physical activity.

\begin{figure*}[]
     \centering
         \includegraphics[width=\textwidth]{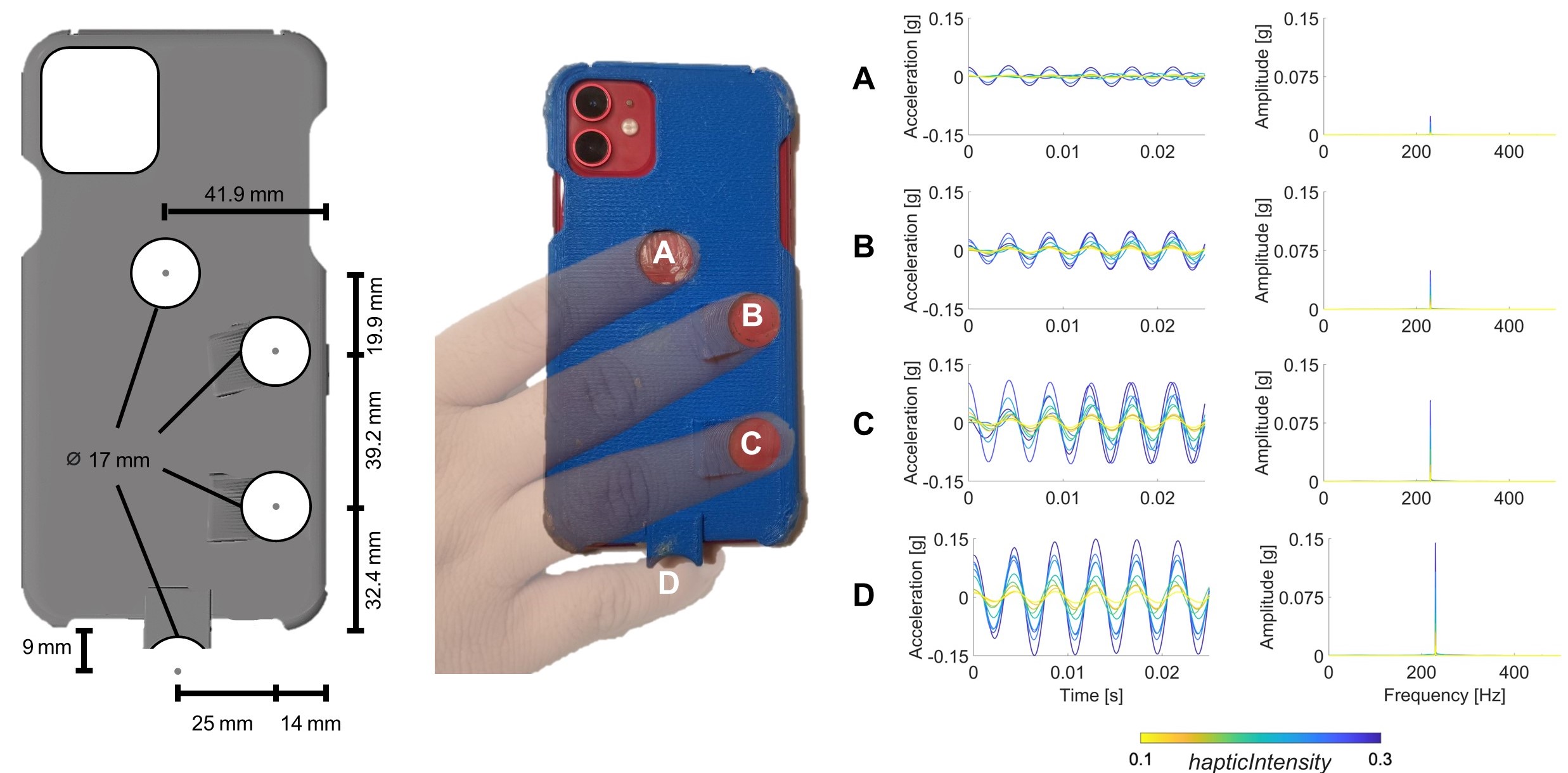}
  \caption{Vibrations were measured at finger contact locations (A: pointer, B: middle, C: ring, D: pinky) from our custom, 3D-printed phone case, which was used to normalize hand placement for participants (left). Vibration amplitude (middle) increased with \textit{hapticIntensity} and was highest at the pinky, where the \textit{taptic engine} is located. Fast Fourier transforms (right) show that peak vibrations occur at 230~Hz.}
  \vspace{-12pt}
  \label{fig: accelerations}
\end{figure*}

\subsection{Smartphone App}
\label{sec:app}

We developed an iOS app (Swift, v5.7.1; XCode v14.1) that executes a vibration-response task and/or a shape-recall task in three modes:

\begin{itemize}
  \item \textit{Vibration-Response Task}: The app delivers vibrations of varying intensities at semi-random time intervals. The user presses a button every time they feel a vibration.
  \item \textit{Shape-Recall Task}: The app flashes a shape image in 3 second intervals. The user presses a button every time they see a shape consecutively shown.
  \item \textit{Vibration-Response and Shape-Recall Task}: The app delivers both modes at the same time and records user responses to both tasks.
\end{itemize}

\subsubsection{Vibration-Response Task}
We created the vibration-response task to measure vibration perception thresholds. 
We used the Core Haptics Framework (Apple Inc.)~\cite{CoreHaptics} to create the smartphone vibrations. \textit{hapticIntensity} and \textit{hapticSharpness} are two unitless parameters in the Core Haptics Framework that can be used to control phone vibrations~\cite{adenekan_phone, Torres2022}. \textit{hapticIntensity} can be set to values between 0-1 to control vibration amplitude, with a larger value resulting in an increased amplitude. \textit{hapticSharpness} can also be set to values between 0-1 and is used to modulate frequency, creating a more ``crisp" signal when increased. In this task, the app uses the method of constant stimuli ~\cite{Gescheider1997Psychophysics} and delivers vibrations with a \textit{hapticIntensity} ranging from 0.100 to 0.300 in 9 equally-spaced 0.125 step increments. \textit{hapticSharpness} is held constant at 1 for all 9 vibration intensities. The duration of each vibration is 0.1 seconds. The 9 vibration intensities are played 10 times each, with each intensity being delivered exactly once before being repeated for a total of 10 sets of 9 vibration intensities. Vibrations are played at random 3-6 second intervals to minimize false positives. Participants are instructed to press the red button (Fig.~\ref{fig:setup}) when they feel a vibration. Vibration response is recorded at the onset of the button press.

\subsubsection{Shape-Recall Task}
We created the shape-recall task to simulate increased cognitive activity for participants. The shape-recall task uses a 1-back version of the n-back task~\cite{kirchner_nback}, commonly used to simulate cognitive load in neuroscience and user interface studies~\cite{Haghighi2020TheEO, Klatzky2007, Ploch2017}. One of 4 distinct shapes (circle, square, triangle, or hexagon) is displayed for 0.5 seconds every 3 seconds. The participant is instructed to press the blue button (Fig.~\ref{fig:setup}) if the current shape matches the shape previously displayed (1-back task). The shape-recall task is grouped into 10 sets of 15 shapes. Within each set, there are exactly 3 randomly distributed shape matches, and therefore 3 times per set when the participant should press the shape response button. This standardizes the quantity and distribution of responses required from the participant. Additionally, this spacing allows for each set of 15 shapes to align with the timing of a set of 9 vibrations. 

\subsubsection{Vibration and Shape-Recall}
In the vibration and shape-recall option, both the shape-recall task and the vibration-response task play concurrently. The participant must press the appropriate response buttons for each task. The time it takes to play a set of 15 shapes equals the time for a set of 9 vibrations. As with the other tasks, 10 sets of each vibration (9 intensities) and shape-recall (15 shapes with 3 shape matches) are played.

\subsection{Smartphone Case}
\label{sec:phone}

In this work, we use an iPhone 11 (Apple Inc.) smartphone. The iPhone 11 has a custom linear resonant actuator at the bottom left side of the phone, commonly known as the \textit{taptic engine}~\cite{iPhone11}. The vibration stimuli are generated by the actuator and then propagate through the phone. Thus, due to damping and other mechanical properties of the phone, the vibration stimuli will vary with respect to the phone contact location. To standardize the vibrations experienced by participants during a user study, we designed a custom\revision{, 3-D printed phone case (PLA, Makerbot)} to guide participants in holding the phone during the user study (Fig.~\ref{fig: accelerations}, top).

To determine the design of the case, we conducted a phone-holding preference pilot study on 10 participants (4 female, 6 male). We painted an off-the-shelf iPhone 11 case with thermochromic paint (Atlanta Chemical Engineering) that changes from black to yellow at 77°F. We attached the painted case to the phone and asked participants to hold the phone comfortably in their right hand. After holding the phone for enough time for the thermochromic paint to change color, participants then placed the phone on a document scanner, and a color scan was taken to capture finger locations shown by the thermochromic paint. The images for each participant were processed using functions within MATLAB's Image Processing Toolbox to convert the images from color to black and white. Using image thresholds, contact locations were converted to white, while the rest of the case was converted to black. We then used these black and white images to create a heat map showing the most common phone contact points. The heat map informed the construction of a phone case with constrained finger placement that would be comfortable for most participants.

The case has three finger holes along the back for the participant’s pointer (Fig.~\ref{fig: accelerations}, A), middle (Fig.~\ref{fig: accelerations}, B), and ring finger (Fig.~\ref{fig: accelerations}, C), along with a divot at the bottom for pinky finger placement (Fig.~\ref{fig: accelerations}, D). The case constrains participants' fingers such that all participants contact the phone in the same places. The middle and ring finger holes have ramps next to them, ensuring participants bend those fingers. The pointer finger is allowed to lay flat against the case, with the finger pad directly contacting the phone. Since the case forces participants to place their fingers on the back and underside of the phone, participants are unable to squeeze the phone. This ensures that participants avoid imparting extraneous forces on the phone since this could potentially change vibration perception.

\subsection{Vibration Characterization}
We used a process similar to Adenekan et al.~\cite{adenekan_phone} to measure smartphone accelerations. An accelerometer (Analog Devices, EVAL-ADXL354CZ, 3-axis, 2~g) and DAQ (National Instruments, NI9220) were used to measure the accelerations at each finger location on the phone case (Fig.~\ref{fig: accelerations}, top) in 0.125 increments of \textit{hapticIntensity} ranging from 0.100 to 0.300 with \textit{hapticSharpness} held constant at 1, matching the vibrations output during the vibration-response task. For each measurement, the phone was placed on a pillow (screen down) and the accelerometer was placed on one of the finger locations~\cite{adenekan_phone}. Data was filtered with a bandpass frequency of 60-500 Hz using MATLAB (Mathworks). 

Figure~\ref{fig: accelerations} shows the filtered acceleration signals and fast Fourier transforms for all \textit{hapticIntensity} values at each location. The highest vibration amplitudes were at the pinky, which is near where the \textit{taptic engine} is located. The vibration amplitudes decrease towards the pointer finger, further from the \textit{taptic engine}. At the pointer finger, amplitudes ranged from 0.003 to 0.03~g. At the pinky, amplitudes ranged from 0.015 to 0.15~g. The frequency peaked at 230~Hz for all values of \textit{hapticIntensity}.

\section{User Study}
\label{sec: user study}
% In order to understand the effects of cognitive and physical load on vibration perception, we created a study design which evaluates each type of distraction individually as well as combined. For the cognitive load,  

We conducted a human subjects user study to better understand the effects of cognitive and physical load on vibration perception. We used the smartphone platform described in the previous section to determine participant vibration thresholds while completing physical and/or cognitive activities.

\subsection{Procedure}

Prior to the study, participants were informed that they had to be right-handed and able to walk for at least 30 minutes. Upon arrival, participants completed a pre-survey form to report their experience with haptic systems in addition to confirming that they met the study requirements and had no other conditions that would impact their performance.

\begin{figure}[]
    \centering
    \includegraphics[width=\columnwidth]{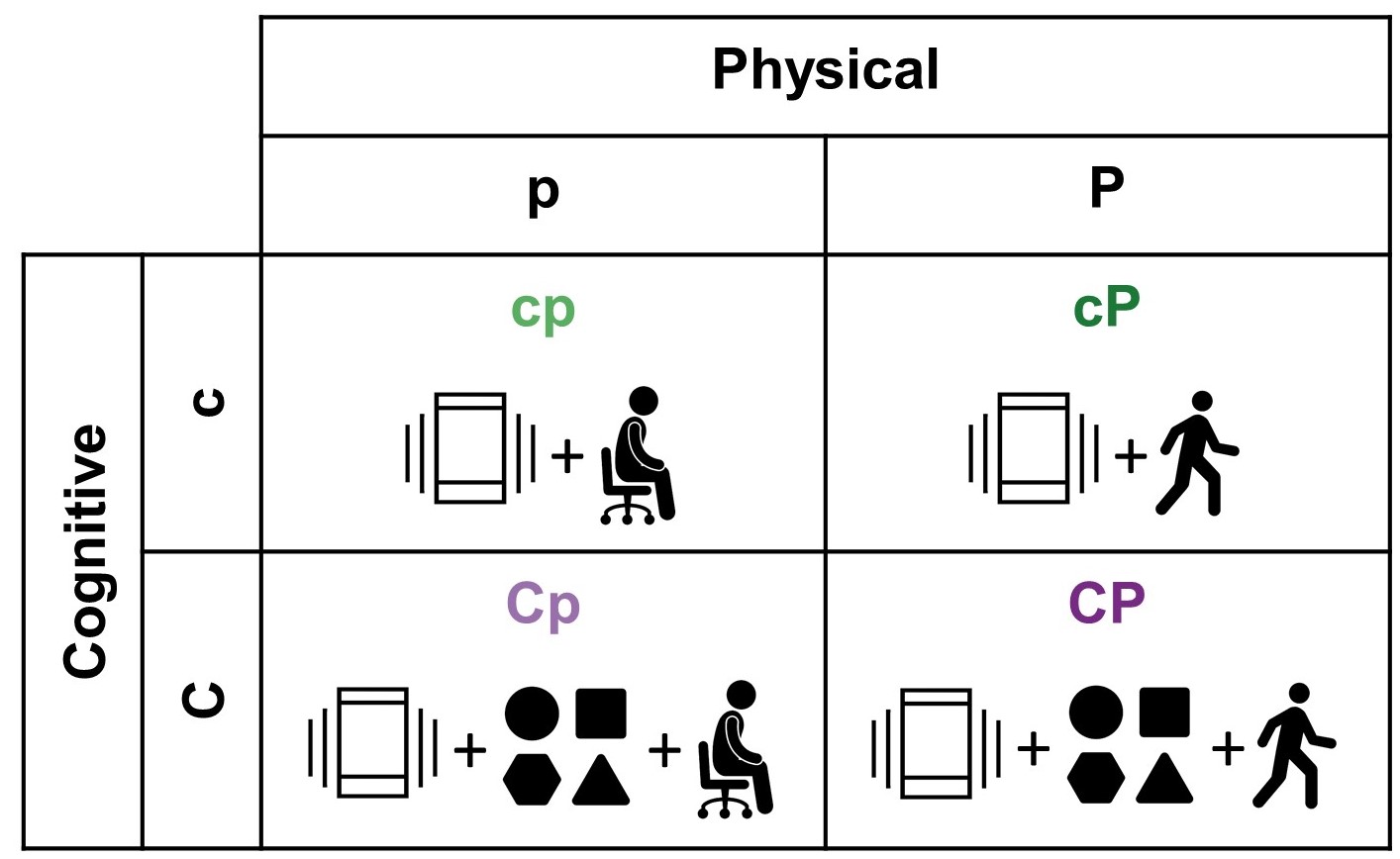}
    \caption{Vibration perception thresholds were measured while the participant had low (c) and high (C) cognitive activity, combined with low (p) or high (P) physical activity. This yielded four test conditions. A shape-recall task introduced cognitive activity (C), while walking on a treadmill introduced physical activity (P).}
    \label{fig: conditions}
    \vspace{-12pt}
\end{figure}

The study was completed in an hour and conducted in five phases (approximately 10 minutes each), corresponding to five different conditions. We measured vibration perception threshold using the vibration-response task during four of these conditions (Fig.~\ref{fig: conditions}), and we added an additional condition for a shape-memory task while sitting to acquaint participants with the shape-memory task format and to establish a baseline to potentially screen out participants who would be unable to complete the task properly. The four conditions in which we measured vibration threshold tested low (c) and high (C) amounts of cognitive activity, combined with either low (p) or high (P) amounts of physical activity. Conditions with low cognitive activity (c) consisted of doing the vibration-response task without the shape-memory task. Conditions with high cognitive activity (C) consisted of the vibration-response task alongside the shape-memory task. Low physical activity (p) meant that participants were seated, and high physical activity (P) meant that participants were walking on a treadmill. After completing the pre-survey, participants were asked adjust the treadmill speed to a comfortable walking pace that they could sustain for 30 minutes. In addition to using the treadmill for the high physical activity condition (P), the treadmill was also used during all sitting tasks to create white noise throughout the study, since walking with headphones presented a safety hazard. 

The first two conditions during the study took place with the subject sitting and consisted of the shape-memory task and the vibration-response task (cp, as shown in Fig.~\ref{fig: conditions}). The order of these two conditions was randomized such that half of the subjects would receive the vibration-response task first, and the other half would receive the shape memory task first. Though internal pilot studies found no effects of ordering or learning, these two conditions were provided first to acquaint users to the system in a safe, seated manner in case they had questions.  

The remaining three conditions (Cp, cP, and CP, as shown in Fig.~\ref{fig: conditions}) test how the remaining combinations of cognitive and physical activity impact vibration perception threshold. The ordering of these three conditions were also pseudo-randomized to test an equal number of participants for each ordering. 

During each condition, participants were instructed to hold the iPhone 11 with the pre-loaded app with their right hand conforming to the finger placement indicated by the phone case (Fig.~\ref{fig: accelerations}). Participants were also instructed to keep their elbow at 90$^{\circ}$ during all conditions (Fig.~\ref{fig:setup}) and to avoid contacting the phone with their left hand. Participants were told to try their best not to prioritize one task over another. After each condition, participants filled out a short task-load index survey ranking mental demand, physical demand, temporal demand, performance, effort, and frustration on a 10-point Likert scale. At the end of the study, participants were asked to rank the conditions in order of difficulty and to provide any additional comments or feedback about the study.

\subsection{Statistical Analysis}
Vibration perception thresholds were calculated by fitting a sigmoid curve to the percentage of vibration detections as a function of \textit{hapticIntensity}. The \textit{hapticIntensity} value on the curve that passes through the point at which 50\% of vibrations were detected, known as the 50\% threshold, was used as the vibration perception threshold~\cite{Gescheider1997Psychophysics}. Any user responses that occurred more than 1.5 seconds after the onset of a vibration cue were discarded as false positives because this was beyond the upper end of haptic response times in literature~\cite{peon_reactiontime, oldmen_reactiontime}. User response time was calculated by taking the difference between the start of the vibration cue and when the participant pressed the vibration response button. A two-way repeated measures ANOVA was used to identify any significant main effects or interaction, followed by post-hoc t-tests with Bonferroni correction. Plots were created in MATLAB (Mathworks) and statistical analysis was conducted in R~\cite{rstudio} using the tidyverse package~\cite{Wickham2019-mu}. 

\subsection{Results}
\label{sec:results}

\begin{figure}[b]
     \centering
         \includegraphics[width=1\columnwidth]{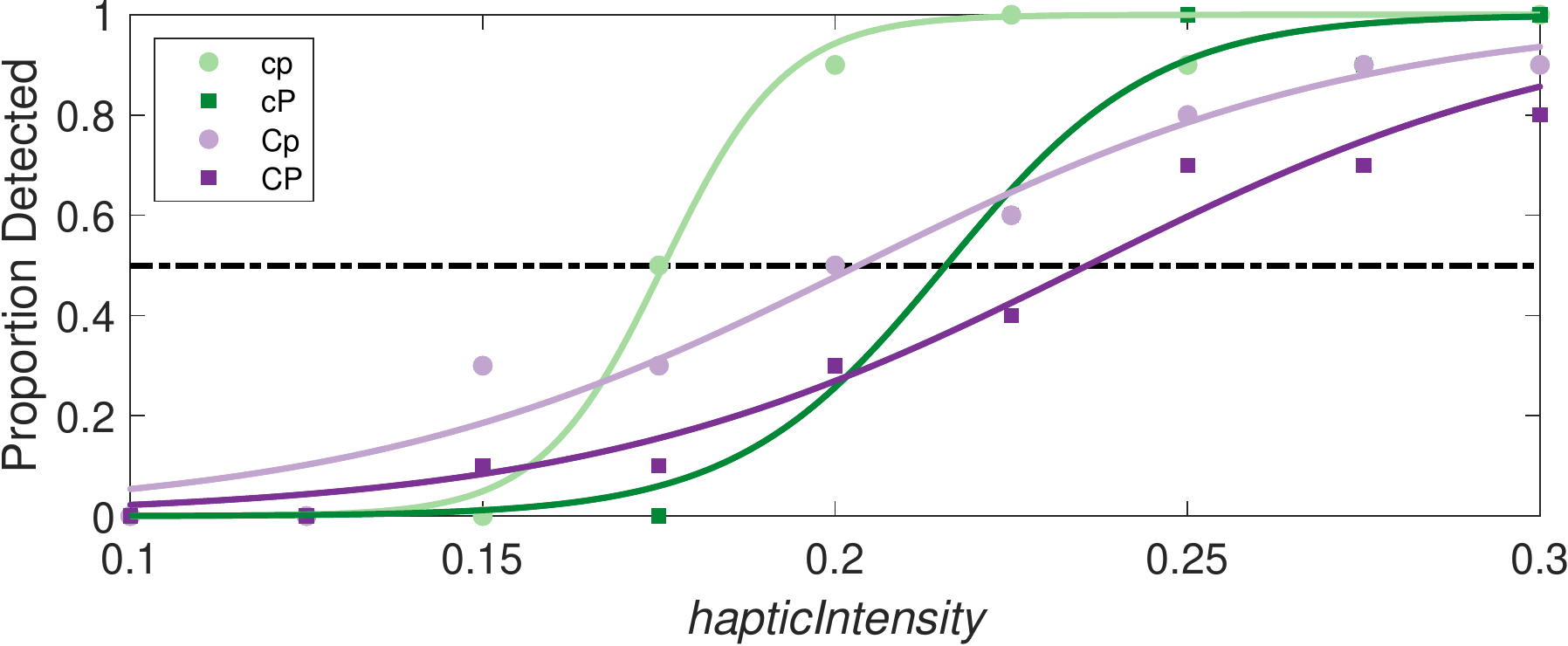}
  \caption{Participant data from a single subject shows the proportion of detected vibrations and the corresponding sigmoid fit for each condition. The \textit{hapticIntensity} corresponding to the intersection of each curve with the 50\% detection point (dashed black line) is the vibration perception threshold. The vibration perception threshold is lower for conditions with low cognitive and physical activity (cp) and moves higher (to the right) for conditions with high cognitive and physical activity (CP). This indicates that increases in cognitive and physical activity impair vibration perception, because a higher \textit{hapticIntensity} is needed for detection.}
  \label{fig: psychometric curve}
\end{figure}

\begin{figure}[]
     \centering
         \includegraphics[width=1\columnwidth]{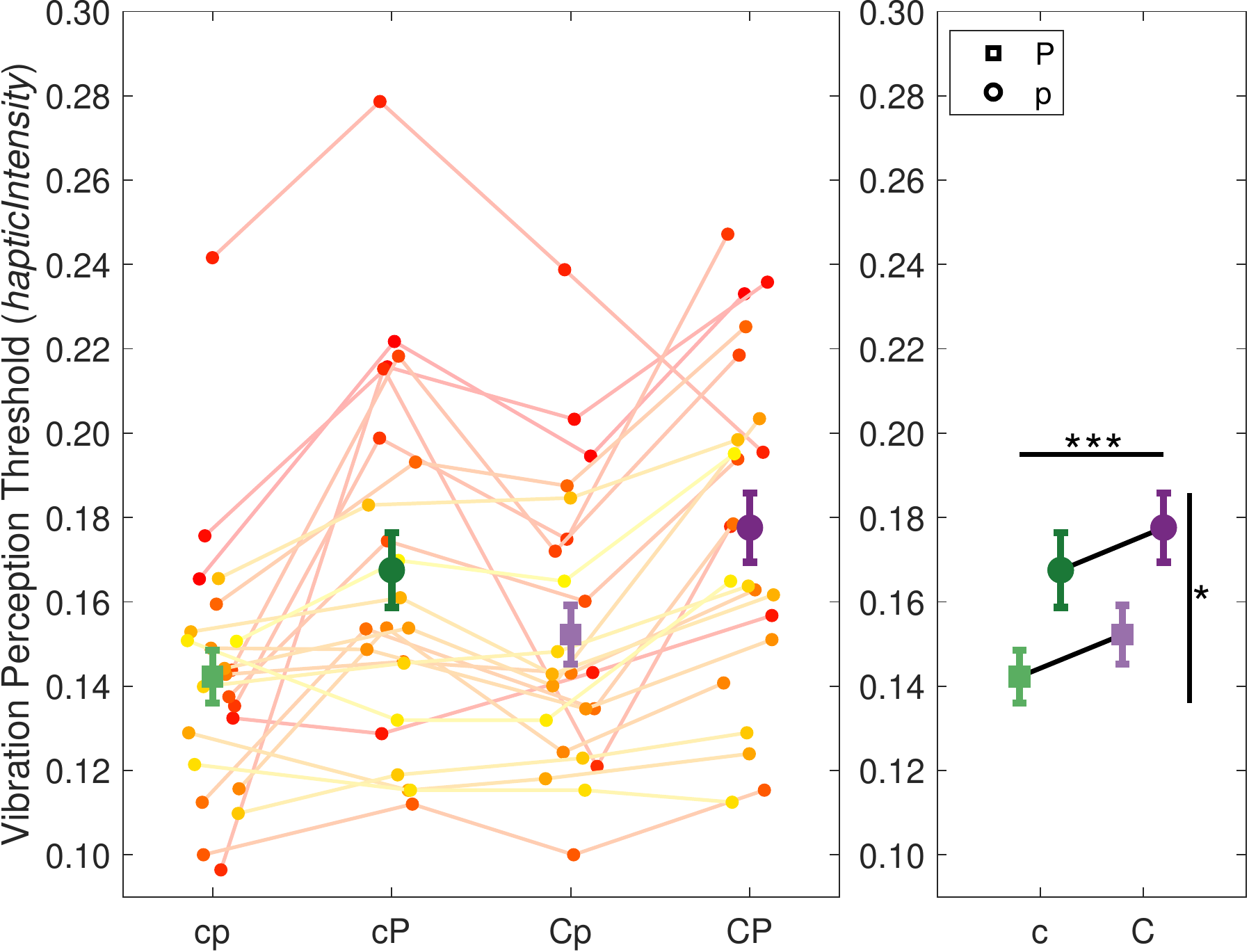}
  \caption{Individual participants (left) and the group average and standard error (right) show that vibration perception threshold significantly increases with high cognitive activity (C) and physical activity (P). Low cognitive (c) and physical (p) activity results in the lowest vibration perception threshold.}
  \label{fig: thresholds}
\end{figure}

\begin{figure}[]
     \centering
         \includegraphics[width=1\columnwidth]{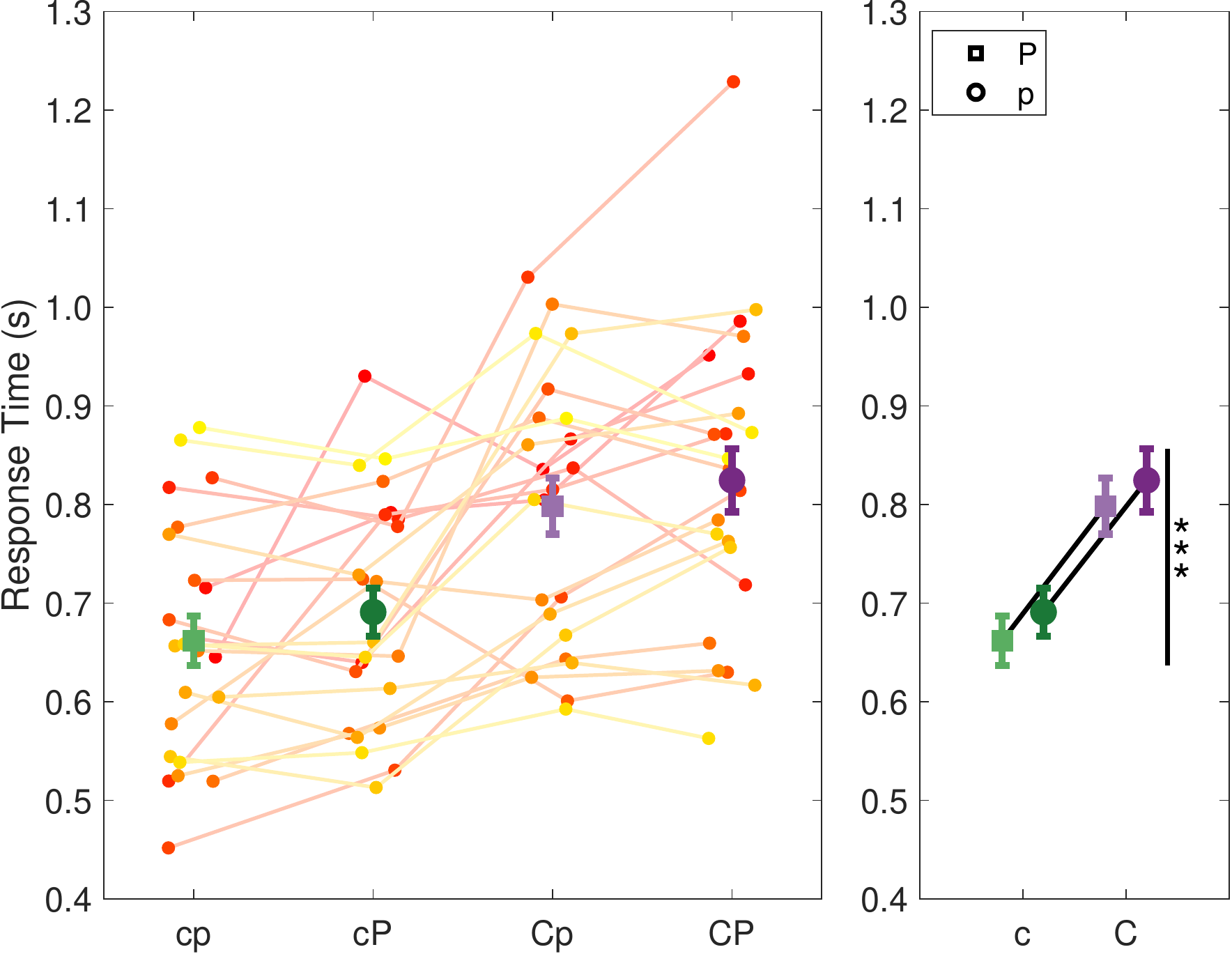}
  \caption{Individual participants (left) and the group average and standard error (right) shows that vibration response time increases significantly with high cognitive activity (C), but not high physical activity (P). Low cognitive (c) and physical (p) activity results in the fastest response time.}
  \label{fig: time}
  \vspace{-12pt}
\end{figure}

We conducted a user study with 24 participants (12 male, 12 female; aged 20-40). Four participants were very familiar with haptic devices and 20 were not. No participants had neurological disorders, injuries to the hand or arm, or other conditions that may have affected their performance in this experiment. This protocol was approved by the Stanford University Institutional Review Board and participants gave informed consent. Participants were compensated with a \$15 gift card for their time. One participant was removed from the corresponding analysis because their data could not be sufficiently fit to the psychometric function. 

\subsubsection{Vibration Perception Threshold}
Figure~\ref{fig: psychometric curve} shows the proportion of detected vibrations with respect to \textit{hapticIntensity} during the four test conditions for a single participant. Aside from the one participant who was removed, vibration perception data from all other participants could be fit adequately ($R_{avg}^{2}=0.959$). Vibration perception thresholds were extracted from individual subject data for analysis. 

Figure~\ref{fig: thresholds} reports the calculated vibration perception thresholds for each participant for each condition as well as the mean threshold and standard error for each condition. The mean vibration perception threshold was lowest ($\textit{hapticIntensity} = 0.142$) in the condition with low cognitive and physical activity (cp) and the highest ($\textit{hapticIntensity} = 0.178$) with high cognitive and physical activity (CP). A two-way, repeated-measures ANOVA with physical activity (low; high) and cognitive activity (low; high) as independent factors, revealed a significant main effect of physical activity ($F(1,22)=25.2$, $p=\num{5.03E-5}$, $\eta_{p}^{2}=0.534$) and cognitive activity ($F(1,22)=7.58$, $p=\num{1.20E-2}$, $\eta_{p}^{2}=0.256$) on vibration perception threshold. These main effects were not qualified by an interaction between cognitive and physical activities ($F(1,22)=\num{3.04E-4}$, $p=0.986$, $\eta_{p}^{2}=\num{1.38E-5}$). %\revision{\sout{Post-hoc pairwise comparisons with Bonferroni correction confirmed that higher cognitive activity ($p=0.004$) and higher physical activity ($p=\num{2.03E-7}$) results in increased vibration perception thresholds. }}

\subsubsection{Response Time}

Figure~\ref{fig: time} presents the response time for each participant for each condition as well as the mean response time and standard error for each condition. Similar to vibration perception thresholds, the mean response time was lowest (0.662~s) in the condition with low cognitive and physical activity (cp) and the highest (0.825~s) with high cognitive and physical activity (CP). A two-way, repeated-measures ANOVA with physical activity (low; high) and cognitive activity (low, high) as independent factors, revealed a main effect of cognitive activity ($F(1,22)=30.5$, $p=\num{1.49E-5}$, $\eta_{p}^{2}=0.136$) and not physical activity ($F(1,22)=3.46$, $p=\num{7.60E-2}$, $\eta_{p}^{2}=0.581$) on response time. These main effects were not qualified by an interaction between cognitive and physical activities ($F(1,22)=0.019$, $p=0.891$, $\eta_{p}^{2}=\num{8.75E-4}$). %\revision{\sout{Post-hoc pairwise comparisons with Bonferroni corrections confirmed that higher physical activity results in increased response times ($p=\num{4.2E-9}$).  }}

\begin{figure}[]
     \centering
         \includegraphics[width=1.0\columnwidth]{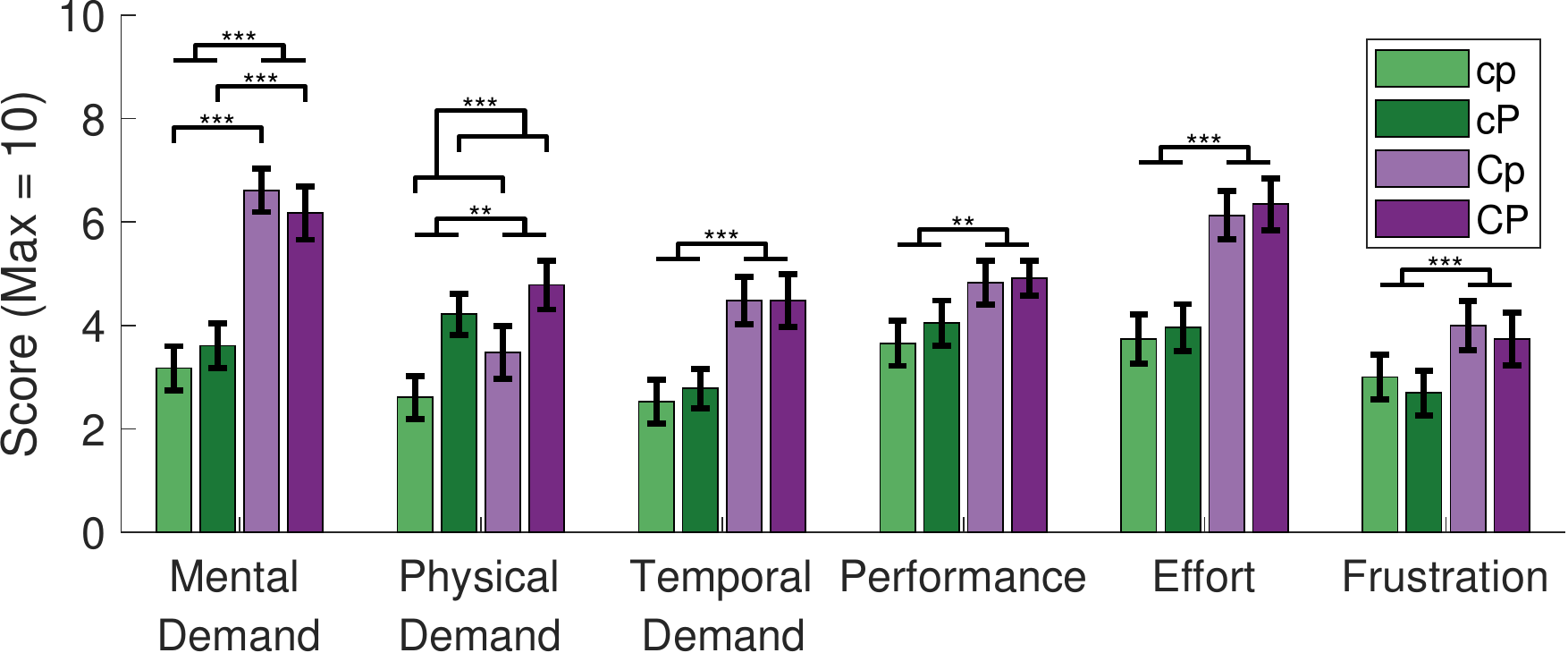}
  %Old caption:
  %\caption{Participants reported lower mental demand with no shape-memory task (c) and lower physical demand when sitting (p). Mental demand increased during the shape-memory task (C) and physical demand increased while walking (P). The shape-memory task (C) resulted in higher frustration and effort, but made participants feel like they performed better.}
  %Improved caption
  \caption{Mean and standard error of participant responses to the task load index surveys. Mental and temporal demand was higher with high cognitive activity (C) as opposed to low cognitive activity (c). Physical demand increased during high physical activity (P) as opposed to low physical activity (p). High cognitive activity (C) resulted in higher frustration and effort, but also made participants feel like they performed better.}
  \label{fig: survey}
  \vspace{-12pt}
\end{figure}

\subsubsection{Survey Responses} Figure~\ref{fig: survey} shows the mean and standard error of the participant responses to the task-load index surveys. Participants reported increased mental and physical demands during our higher cognitive and physical conditions, respectively. Although participants report more effort, frustration, and temporal demand during the conditions with higher cognitive activity, participants thought that they performed better in these conditions.

\underline{Mental Load}: A two-way, repeated-measures ANOVA revealed a main effect of cognitive activity ($F(1,22)=45.3$, $p=\num{9.14E-7}$, $\eta_{p}^{2}=0.673$) and not physical activity ($F(1,22)=\num{2.73E-30}$, $p=1.00$, $\eta_{p}^{2}=\num{1.24E-31}$) on participant-reported mental load. These main effects were qualified by an interaction between cognitive and physical activities ($F(1, 22)=\num{4.52}$, $p=0.045$, $\eta_{p}^{2}=0.171$). Post-hoc comparison with Bonforroni corrections confirmed that the simple main effect of cognitive load was significant at high ($F(1,22)=47.5$, $p=\num{1.28E-6}$, $\eta_{p}^{2}=0.431$) and low levels of physical activity ($F(1,22)=28.3$, $p=\num{4.88E-5}$, $\eta_{p}^{2}=0.245$). Pairwise comparisons show that participant-reported mental load was significantly different between low and high cognitive activity for low physical activity ($p=\num{6.38E-7}$) and high physical activity ($p=\num{2.44E-5}$). 

\underline{Physical Load}: A two-way, repeated-measures ANOVA revealed a significant main effect of physical activity ($F(1,22)=6.29$, $p=0.020$, $\eta_{p}^{2}=0.222$) and cognitive activity ($F(1,22)=12.6$, $p=0.002$, $\eta_{p}^{2}=0.363$) on participant-reported physical load. These main effects were not qualified by an interaction between cognitive and physical activities ($F(1, 22)=0.566$, $p=0.460$, $\eta_{p}^{2}=0.025$). %\revision{\sout{Post-hoc pairwise comparisons with Bonferroni correction confirmed that higher cognitive activity ($p=0.006$) and higher physical activity ($p=\num{4.23E-5}$) results in higher participant-reported physical load. }}

\underline{Temporal Demand, Performance, Effort, and Frustration}: Two-way, repeated measures ANOVAs showed that %\revision{\sout{only}} 
cognitive activity had a main effect on participant-reported temporal demand \revision{($F(1,22)=17.3$, $p=\num{4.04E-4}$, $\eta_{p}^{2}=0.441$)}, performance \revision{($F(1,22)=10.8$, $p=0.003$, $\eta_{p}^{2}=0.330$)}, effort \revision{($F(1,22)=39.1$, $p=\num{2.71E-6}$, $\eta_{p}^{2}=0.640$)}, and frustration \revision{($F(1,22)=8.84$, $p=0.007$, $\eta_{p}^{2}=0.287$)}%\revision{\sout{($p<0.05$)}}
. 

\revision{Two-way, repeated measures ANOVAs showed that physical activity did not have a main effect on participant-reported temporal demand ($F(1,22)=0.489$, $p=0.492$, $\eta_{p}^{2}=0.022$), performance ($F(1,22)=1.05$, $p=0.316$, $\eta_{p}^{2}=0.046$), effort ($F(1,22)=1.12$, $p=0.302$, $\eta_{p}^{2}=0.048$), or frustration ($F(1,22)=1.31$, $p=0.265$, $\eta_{p}^{2}=0.056$). }

%\revision{\sout{Post-hoc analysis for all of these participant-reported measures confirmed that higher levels of cognitive activity resulted in increases in participant-reported temporal demand, performance, effort, and frustration ($p<0.05$). }}

\section{Discussion}
\label{sec:discussion}

This study found that cognitive and physical activities result in increased vibration perception thresholds and that cognitive activity results in slower response times to these vibrations. The average vibration perception threshold range in our study (\textit{hapticIntensity} of 0.142 to 0.178) was lower than that found in a study that used a smartphone for absolute threshold clinical testing using the staircase method~\cite{adenekan_phone}. These differences could be due to differences in the phone model and psychophysical method used for measuring vibration threshold. Similar to another study using smartphones for clinical testing, our study found that the 50\% vibration perception threshold fell between a \textit{hapticIntensity} of 0.100 to 0.300, with most subject perception saturating above a \textit{hapticIntensity} of 0.300~\cite{Torres2022}. Our method also produced better psychometric fits than Torres et al.~\cite{Torres2022}, indicating that our range and finer resolution should be used for future studies on smartphone vibration perception thresholds. \revision{Similar to other studies, we found that there was also impaired vibration perception from cognitive and physical activity~\cite{karuei_acrossbodywalking, ANGEL1988cutaneoussensitivity, chapquouo2018restandwalking, Chapwouo2018, Klatzky2007, yildiz_activeandpassive_2015, post_zompa_chapman_1994, Haghighi2020TheEO}.}

The response time to vibrations in our study (averages from 0.662 to 0.825~s) were similar to other setups (0.150 to 0.900~s), with minor differences likely due to the amount of time required to physically move to respond~\cite{peon_reactiontime, oldmen_reactiontime}. Our study showed that response time is primarily affected by cognitive activity and not physical activity\revision{, similar to other studies on road distractions~\cite{daddario_donmez_2019}}. Although the button was placed in the same location for all test conditions, participants may have needed extra time when multiple buttons are present to further process which button needed to be pressed. 

The accelerometer measurements showed that vibration amplitudes can be well-controlled in 0.125 \textit{hapticIntensity} increments, and has a peak frequency at 230~Hz. Our phone case, created by identifying a common holding position for the iPhone 11, can be used and adapted for other studies to control for hand positioning, as vibration accelerations vary throughout the phone.

\section{Conclusion}
\label{sec:conclusion}
We found that cognitive and physical activities increase vibration perception thresholds and that cognitive activity results in slower response times. Our results show that when engineers and haptics researchers design devices, they should account for changes in the environment, as cognitive and physical activity both contribute to changes in perception. 

We also develop and characterize a smartphone platform that can be used for future experiments in haptics that use a smartphone. Our phone case used to control for differences in holding positions to normalize vibrations across users and our acceleration characterizations can be used to better understand the parameters in the Apple Core Haptics Library. 

In the future, we will develop more studies using smartphones that can be widely distributed to diverse populations. We are also working on future studies that use the phone to test for clinical sensory tests and to examine the effect of haptic intensity on reaction times, among many others. 

\section{Acknowledgements}

We would like to thank E. Chase, E. Childs, and N. Agharese for insightful discussions on data analysis.

\begin{comment}

\begin{figure}[]
     \centering
         \includegraphics[width=\columnwidth]{Figures/fig3.pdf}
     \vspace{-0.25in}
  \caption{caption}
  \label{plot}
\end{figure}
\end{comment}

%\bibliographystyle{IEEEtran}

\bibliographystyle{IEEEtran}
\bibliography{distraction.bib}

\begin{thebibliography}{10}
\providecommand{\url}[1]{#1}
\csname url@samestyle\endcsname
\providecommand{\newblock}{\relax}
\providecommand{\bibinfo}[2]{#2}
\providecommand{\BIBentrySTDinterwordspacing}{\spaceskip=0pt\relax}
\providecommand{\BIBentryALTinterwordstretchfactor}{4}
\providecommand{\BIBentryALTinterwordspacing}{\spaceskip=\fontdimen2\font plus
\BIBentryALTinterwordstretchfactor\fontdimen3\font minus
  \fontdimen4\font\relax}
\providecommand{\BIBforeignlanguage}[2]{{%
\expandafter\ifx\csname l@#1\endcsname\relax
\typeout{** WARNING: IEEEtran.bst: No hyphenation pattern has been}%
\typeout{** loaded for the language `#1'. Using the pattern for}%
\typeout{** the default language instead.}%
\else
\language=\csname l@#1\endcsname
\fi
#2}}
\providecommand{\BIBdecl}{\relax}
\BIBdecl

\bibitem{Blum2019OutsideLab}
J.~R. Blum \emph{et~al.}, ``{Getting Your Hands Dirty Outside the Lab: A
  Practical Primer for Conducting Wearable Vibrotactile Haptics Research},''
  \emph{IEEE Transactions on Haptics}, vol.~12, no.~3, pp. 232--246, 2019.

\bibitem{post_zompa_chapman_1994}
L.~Post, I.~Zompa, and C.~Chapman, ``Perception of vibrotactile stimuli during
  motor activity in human subjects,'' \emph{Experimental Brain Research}, vol.
  100, no.~1, 1994.

\bibitem{yildiz_activeandpassive_2015}
M.~Z. Yildiz, I.~Karakuş, F.~Ozkan, and B.~Guclu, ``Effects of passive and
  active movement on vibrotactile detection thresholds of the pacinian channel
  and forward masking,'' \emph{Somatosensory and Motor Research}, vol.~32, pp.
  1--11, 10 2015.

\bibitem{ANGEL1988cutaneoussensitivity}
R.~W. Angel and R.~C. Malenka, ``Velocity-dependent suppression of cutaneous
  sensitivity during movement,'' \emph{Experimental Neurology}, vol.~77, no.~2,
  pp. 266--274, 1982.

\bibitem{karuei_acrossbodywalking}
I.~Karuei, K.~E. MacLean, Z.~Foley-Fisher, R.~MacKenzie, S.~Koch, and
  M.~El-Zohairy, ``Detecting vibrations across the body in mobile contexts,''
  in \emph{Proceedings of the SIGCHI Conference on Human Factors in Computing
  Systems}.\hskip 1em plus 0.5em minus 0.4em\relax Association for Computing
  Machinery, 2011, p. 3267–3276.

\bibitem{chapquouo2018restandwalking}
L.~D. Chapwouo~Tchakouté, L.~Tremblay, and B.-A.~J. Menelas, ``Response time
  to a vibrotactile stimulus presented on the foot at rest and during walking
  on different surfaces,'' \emph{Sensors}, vol.~18, no.~7, 2018.

\bibitem{Chapwouo2018}
L.~Chapwouo~T. and B.~Menelas, ``Impact of auditory distractions on haptic
  messages presented under the foot,'' \emph{International Conference on Human
  Computer Interaction Theory and Applications}, pp. 55--63, 2018.

\bibitem{Klatzky2007}
R.~Klatzky, N.~Giudice, J.~Marston, J.~Tietz, R.~Golledge, and J.~Loomis, ``An
  n-back task using vibrotactile stimulation with comparison to an auditory
  analogue,'' \emph{Behavior Research Methods}, vol.~40, pp. 367--72, 2007.

\bibitem{Haghighi2020TheEO}
N.~Haghighi, N.~A. Vladis, Y.~Liu, and A.~Satyanarayan, ``The effectiveness of
  haptic properties under cognitive load: An exploratory study,'' \emph{ArXiv},
  2020.

\bibitem{Ploch2017}
C.~J. Ploch, J.~H. Bae, C.~C. Ploch, W.~Ju, and M.~R. Cutkosky, ``Comparing
  haptic and audio navigation cues on the road for distracted drivers with a
  skin stretch steering wheel,'' in \emph{IEEE World Haptics Conference}, 2017,
  pp. 448--453.

\bibitem{shah2018dualtaskreaching}
V.~Shah, N.~Risi, G.~Ballardini, L.~Mrotek, M.~Casadio, and R.~Scheidt,
  ``Effect of dual tasking on vibrotactile feedback guided reaching – a pilot
  study,'' \emph{Haptics}, vol. 10893, pp. 3--14, 06 2018.

\bibitem{daddario_donmez_2019}
P.~D’Addario and B.~Donmez, ``The effect of cognitive distraction on
  perception-response time to unexpected abrupt and gradually onset roadway
  hazards,'' \emph{Accident Analysis and Prevention}, vol. 127, p. 177–185,
  2019.

\bibitem{yoshida_3dof_worldhaptics}
K.~T. Yoshida, C.~M. Nunez, S.~R. Williams, A.~M. Okamura, and M.~Luo, ``3-dof
  wearable, pneumatic haptic device to deliver normal, shear, vibration, and
  torsion feedback,'' in \emph{2019 IEEE World Haptics Conference}, 2019, pp.
  97--102.

\bibitem{socialtouch}
M.~Salvato \emph{et~al.}, ``Data-driven sparse skin stimulation can convey
  social touch information to humans,'' \emph{IEEE Transactions on Haptics},
  vol.~15, no.~2, pp. 392--404, 2022.

\bibitem{adenekan_phone}
R.~A.~G. Adenekan, A.~J. Lowber, B.~Huerta, A.~M. Okamura, K.~T. Yoshida, and
  C.~M. Nunez, ``{Feasibility of Smartphone Vibrations as a Sensory Diagnostic
  Tool},'' \emph{Haptics: Science, Technology, Applications}, pp. 337--339,
  2022.

\bibitem{Torres2022}
\BIBentryALTinterwordspacing
W.~Torres, M.~Abbott, Y.~Wang, and H.~Stuart, ``Cutaneous perception
  identification using smartphone haptic feedback,'' \emph{TechRxiv}, 2022.
  [Online]. Available: \url{https://doi.org/10.36227/techrxiv.20493039.v1}
\BIBentrySTDinterwordspacing

\bibitem{CoreHaptics}
{Apple, Inc.}, \emph{Core Haptics Developer Documentation}, [Accessed:
  12-Dec-2022]. Available: developer.apple.com/documentation/corehaptics.

\bibitem{Gescheider1997Psychophysics}
G.~A. Gescheider, \emph{Psychophysics: The Fundamentals}.\hskip 1em plus 0.5em
  minus 0.4em\relax Lawrence Erlbaum Associates, Inc., 1997.

\bibitem{kirchner_nback}
W.~K. Kirchner, ``Age differences in short-term retention of rapidly changing
  information.'' \emph{Journal of Experimental Psychology}, vol. 55 4, pp.
  352--8, 1958.

\bibitem{iPhone11}
J.~Suovanen. (2021) {iPhone 11 Taptic Engine Replacement}. iFixit. [Accessed:
  19-Jan-2022]. Available: https://www.ifixit.com
  /Guide/iPhone+11+Taptic+Engine+Replacement/130323.

\bibitem{peon_reactiontime}
A.~R. Peon and D.~Prattichizzo, ``Reaction times to constraint violation in
  haptics: comparing vibration, visual and audio stimuli,'' in \emph{World
  Haptics Conference}, 2013, pp. 657--661.

\bibitem{oldmen_reactiontime}
P.~Era, J.~Jokela, and E.~Heikkinen, ``{Reaction and Movement Times in Men of
  Different Ages: A Population Study},'' \emph{Perceptual and Motor Skills},
  vol.~63, no.~1, pp. 111--130, 1986.

\bibitem{rstudio}
{RStudio Team}, \emph{RStudio: Integrated Development Environment for R},
  RStudio, PBC., Boston, MA, 2020.

\bibitem{Wickham2019-mu}
H.~Wickham \emph{et~al.}, ``Welcome to the tidyverse,'' \emph{Journal of Open
  Source Software}, vol.~4, no.~43, p. 1686, 2019.

\end{thebibliography}

%\newpage
\vspace{-34pt}

\begin{IEEEbiography}[{\includegraphics[width=1in,height=1.25in,clip,keepaspectratio]{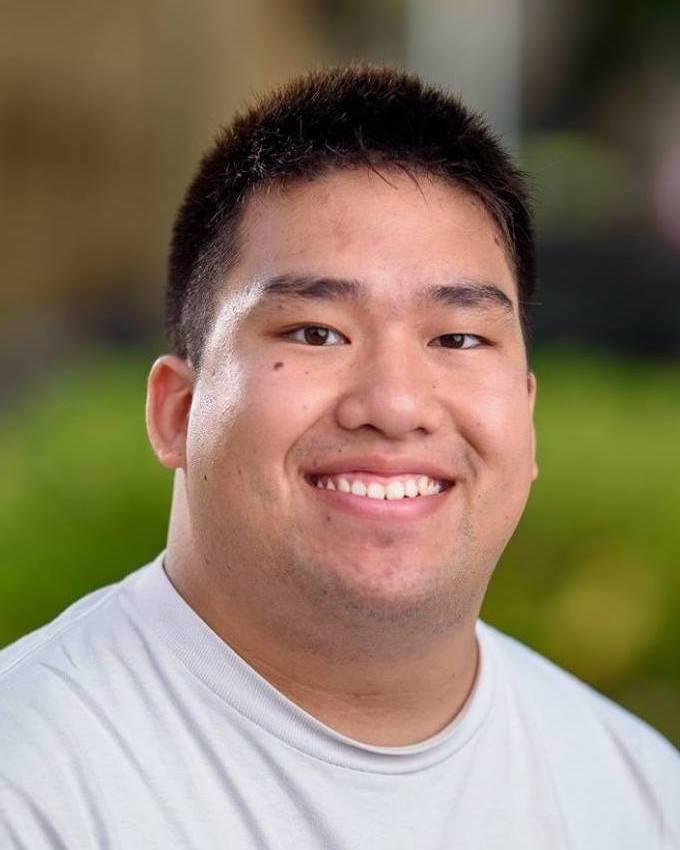}}]{Kyle T.\ Yoshida}
received the S.B. degree in bioengineering and a minor in African studies from Harvard University, Cambridge, MA, USA, in 2018 and the M.S. degree in mechanical engineering from Stanford University, Stanford, CA, USA, in 2020. He is a National Science Foundation Graduate Research Fellow and Stanford Graduate Fellow. His research interests include mobile and wearable haptics.
\end{IEEEbiography}

\vspace{-30pt}

\begin{IEEEbiography}[{\includegraphics[width=1in,height=1.25in,clip,keepaspectratio]{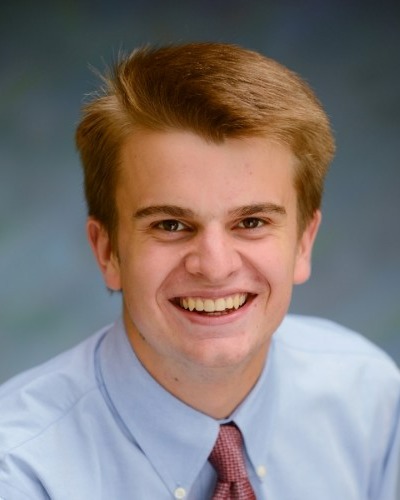}}]{Joel X.\ Kiernan}
is pursuing the B.S. degree in mechanical engineering from Stanford University, Stanford, CA, USA. His research interests include mobile haptics and human perception.
\end{IEEEbiography}

\vspace{50pt}

\begin{IEEEbiography}[{\includegraphics[width=1in,height=1.25in,clip,keepaspectratio]{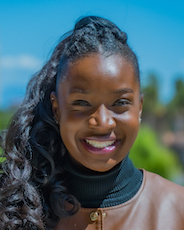}}]{Rachel A.\ G.\ Adenekan}
received the B.S. degree in mechanical engineering and a minor in music from the Massachusetts Institute of Technology, Cambridge, MA, USA, in 2017 and the M.S. degree in mechanical engineering from Stanford University, Stanford, CA, USA, in 2019. She is a National Science Foundation Graduate Research Fellow and a Stanford Graduate Fellow (Medtronic Foundations Fellow). Her research interests include mobile haptics for health applications, digital health technologies, and biomechanics.
\end{IEEEbiography}

\vspace{-30pt}

\begin{IEEEbiography}[{\includegraphics[width=1in,height=1.25in,clip,keepaspectratio]{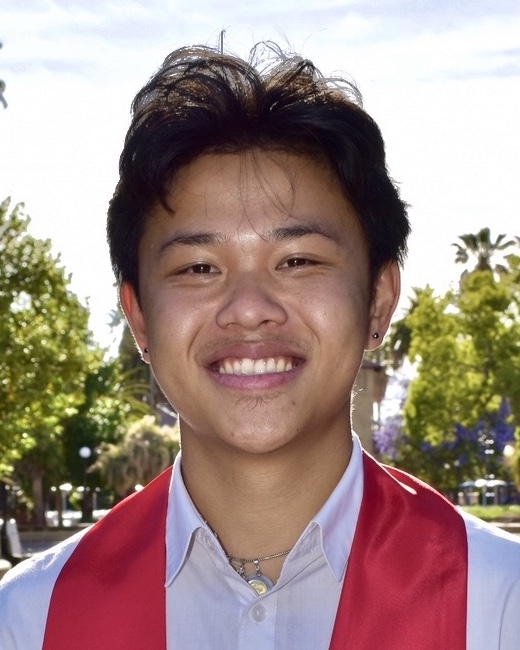}}]{Steven H.\ Trinh}
is pursuing the B.S. degree in engineering product design and the M.S. degree in mechanical engineering from Stanford University, Stanford, CA, USA. His research interests include mobile and wearable haptics and human-computer interaction.
\end{IEEEbiography}

\vspace{-30pt}

\begin{IEEEbiography}[{\includegraphics[width=1in,height=1.25in,clip,keepaspectratio]{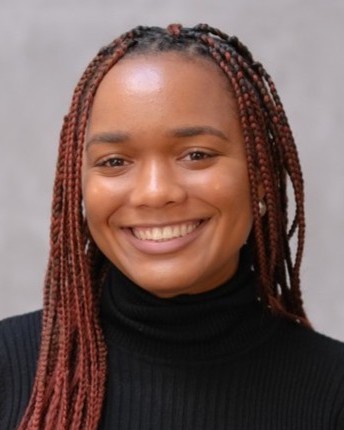}}]{Alexis J.\ Lowber}
received the B.S. degree in computer science and a minor in Spanish from Stanford University, Stanford, CA, USA, in 2021. She will receive her M.S. degree in computer science from Stanford University in 2023 and will begin working as a software engineer at Google, Sunnyvale, CA, USA. Her research interests include mobile and wearable haptics and medical devices.
\end{IEEEbiography}

\vspace{-30pt}

\begin{IEEEbiography}[{\includegraphics[width=1in,height=1.25in,clip,keepaspectratio]{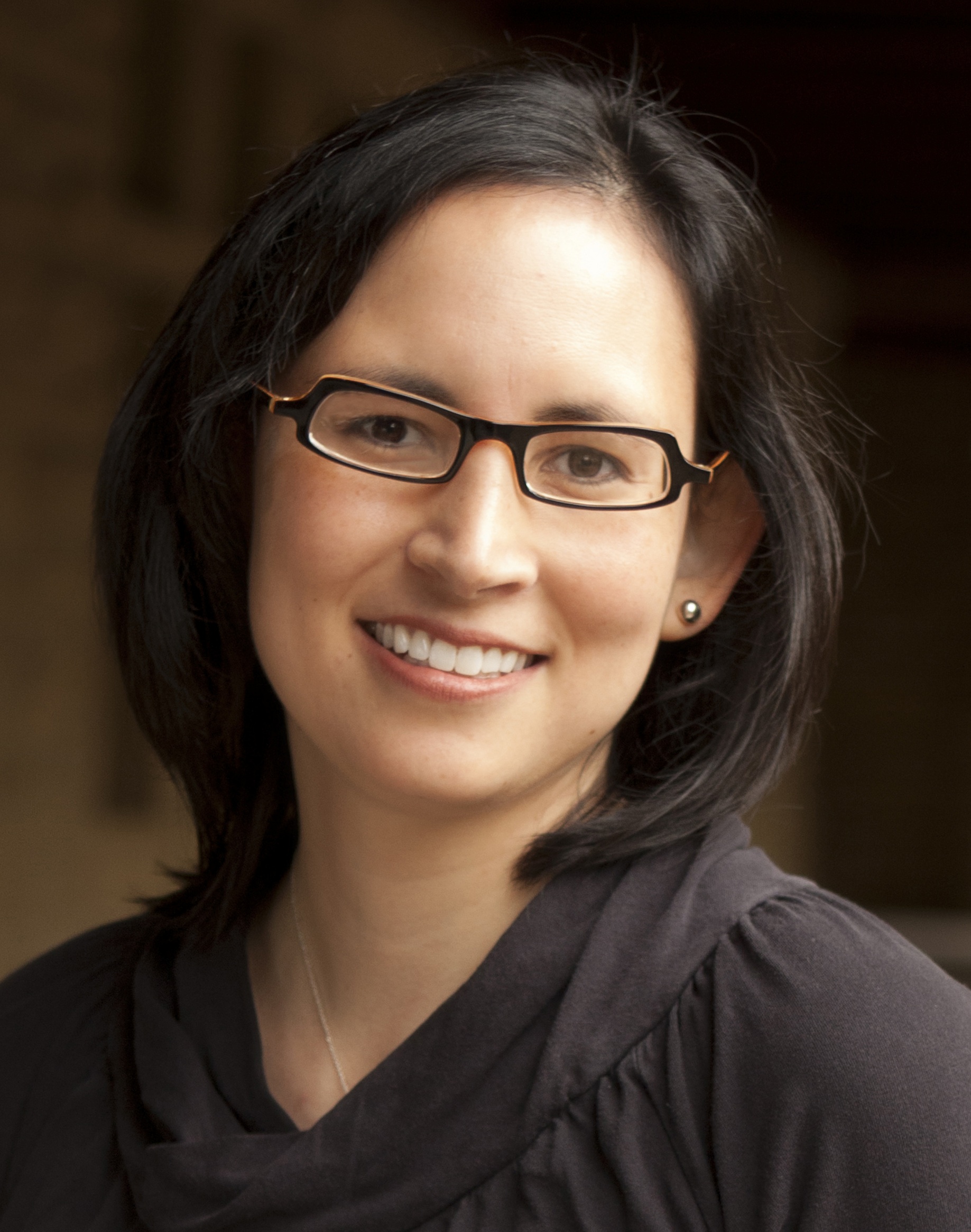}}]{Allison M.\ Okamura} (Fellow, IEEE) received the B.S. degree from the University of California, Berkeley, Berkeley, CA, USA, in 1994, and the M.S. and Ph.D. degrees from Stanford University, Stanford, CA, USA, in 1996 and 2000, respectively, all in mechanical engineering. She is currently the Richard W. Weiland Professor in the School of Engineering and Professor of Mechanical Engineering with Stanford University. Her research interests include haptics, teleoperation, medical robotics, virtual environments and simulation, neuromechanics and rehabilitation, prosthetics, and engineering education. Prof.\ Okamura's awards include the 2020 IEEE Engineering in Medicine and Biology Society Technical Achievement Award and the 2019 IEEE Robotics and Automation Society Distinguished Service Award. She was the Editor-in-Chief of the \emph{IEEE Robotics and Automation Letters} from 2018-2021.
\end{IEEEbiography}

%\begin{IEEEbiography}[{\includegraphics[width=1in,height=1.25in,clip,keepaspectratio]{headshots/Nunez.jpg}}]{Cara M.\ Nunez} (Member, IEEE) received the B.S. degree in biomedical engineering and the B.A. degree in Spanish from the University of Rhode Island, Kingston, RI, USA, in 2016 and the M.S. degree in mechanical engineering and the Ph.D. degree in bioengineering from Stanford University, Stanford, CA, USA, in 2018 and 2021, respectively. She is currently an Assistant Professor in the Sibley School of Mechanical and Aerospace Engineering at Cornell University, Ithaca, NY, USA. Her research interests include haptics and robotics for medical applications, human-machine interaction, augmented and virtual reality, and STEM education. Prof.\ Nunez was the Student Activities Committee Chair from 2020-2022 and is currently the Student Activities Committee Senior Chair and an Associate Vice President for the Media Services Board for the IEEE Robotics and Automation Society. 

%\end{IEEEbiography}

% %%% include if space
\vspace{-30pt}

\begin{IEEEbiography}[{\includegraphics[width=1in,height=1.25in,clip,keepaspectratio]{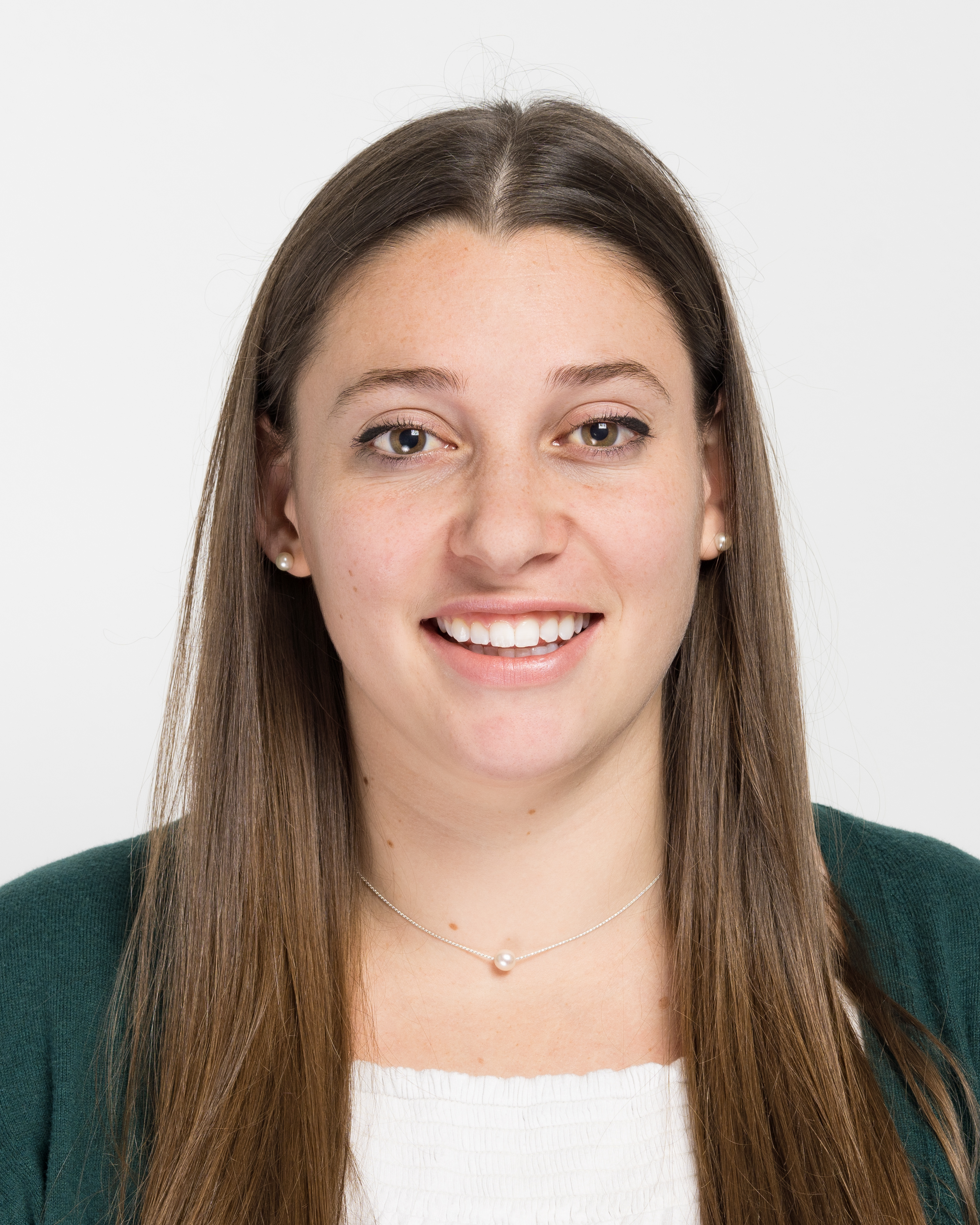}}]{Cara M.\ Nunez} (Member, IEEE) received the B.S. degree in biomedical engineering and the B.A. degree in Spanish from the University of Rhode Island, Kingston, RI, USA, in 2016 and the M.S. degree in mechanical engineering and the Ph.D. degree in bioengineering from Stanford University, Stanford, CA, USA, in 2018 and 2021, respectively.

%From 2019 to 2020, she was a DAAD Graduate Research Fellow at the Max Planck Institute for Intelligent Systems, Stuttgart, Germany. In 2021, she was a Robotics Research Intern at the Honda Research Institute, USA, San Jose, CA, USA. From 2021 to 2023, she was a Postdoctoral Research Fellow at the Harvard John A. Paulson School of Engineering and Applied Sciences, Cambridge, MA, USA. 
She is currently an Assistant Professor in the Sibley School of Mechanical and Aerospace Engineering at Cornell University, Ithaca, NY, USA. She was formerly a DAAD Graduate Research Fellow at the Max Planck Institute for Intelligent Systems, Stuttgart, Germany, a Robotics Research Intern at the Honda Research Institute, San Jose, CA, USA, and a Postdoctoral Research Fellow at the Harvard John A. Paulson School of Engineering and Applied Sciences, Cambridge, MA, USA. Her research interests include haptics and robotics for medical applications, human-machine interaction, augmented and virtual reality, and STEM education. %Her research interests include haptics, robotics, wearable devices, and human perception for medical applications, human-robot interaction, virtual reality, and STEM education. %Her research interests include haptic perception, cutaneous feedback, and wearable devices for medical applications, human-robot interaction, virtual reality, and STEM education.

Prof.\ Nunez was the Student Activities Committee Chair from 2020-2022 and is currently the Student Activities Committee Senior Chair and an Associate Vice President for the Media Services Board for the IEEE Robotics and Automation Society.

 \end{IEEEbiography}


\vfill

\end{document}